\documentclass[]{elsart3}

\ifx\pdftexversion\undefined
  \usepackage[dvips]{graphicx}
\else
  \usepackage[pdftex]{graphicx}
\fi

\usepackage{amsfonts}
\usepackage{amssymb}

\input {isolatin1.sty}
\begin{document}

\def\chargedensity{\rho}
\def\cal{}

\def\E{{\bf E}}

\def\Q{{\bf Q}}

\def\J{{\bf J}}

\def\B{{\bf B}}

\def\D{{\bf D}}

\def\r{{\bf r}} \def\v{{\bf v}} 

\def\dV{{\; \rm d^3}{\bf r}}

\def\curl{{{ \rm curl}\; }} 

\def\grad{{{ \rm grad}\; }}

\def\div{{{\rm div}\; }} 

\def\p{{\bf p}} 

\def\U{{\mathcal U}} 

\def\dv{{\rm d^3}{\bf r}} 

\def\Z{{\mathcal Z}}
\begin{frontmatter}
  
  \title{Auxiliary field simulation and Coulomb's Law} \author[l1]{A.C.
    Maggs}, \author[l2]{J. Rottler } \address[l1]{ Laboratoire de
    Physico-Chimie Th\'eorique, UMR CNRS-ESPCI 7083, 10 rue Vauquelin,
    F-75231 Paris Cedex 05, France.  }  \address[l2]{ Princeton
    Institute for the Science and Technology of Materials (PRISM),
    Princeton University, Princeton, NJ 08544, USA.  }

\begin{abstract}
  We review a family of local algorithms that permit the simulation of
  charged particles with purely local dynamics. Molecular dynamics
  formulations lead to discretizations similar to those of ``particle
  in cell'' methods in plasma physics. We show how to formulate a
  local Monte-Carlo algorithm in the presence of the long ranged
  Coulomb interaction. We compare the efficiencies of our molecular
  dynamics and Monte-Carlo codes.
\end{abstract}
\end{frontmatter}
\maketitle
\section{Introduction}
Computer modeling of charged systems is demanding due to the range of
the Coulomb interaction \cite{schlick}. The direct evaluation of the
Coulomb sum for $N$ particles, $U_c=\sum_{i<j}e_ie_j/4\pi\epsilon_0
r_{ij}$, requires computation of the separations $r_{ij}$ between all
pairs of particles, which implies $O(N^2)$ operations are needed per
sweep or time step.  Most large scale codes in use at the moment find
the electrostatic energy and forces using the fast Fourier transform
\cite{darden} after interpolating charges to a mesh. For an ensemble
of charges interpolated to $O(M)$ nodes of a lattice this gives the
electrostatic energy in a time which scales as $O(N+M \ln{M})$ on a
mono-processor machine.

It is rather surprising that methods that are based on solution of the
Poisson equation are so dominant in mesh based molecular dynamics.  In
plasma physics, one may use a global solver or alternatively one
integrates the full {\sl local} Maxwell equations for the
electromagnetic field. We shall show in this article that a similar
approach based on Maxwell's equations also gives good results in
molecular dynamics; Coulomb's law is implied by Maxwell's equations in
the quasi-static limit. The advantage of this approach is that
parallel implementations which exhibit $O(N+M)$ scaling are rather easy
to write.

After abstracting the part of Maxwell's equations which is essential
for Coulomb's law we introduce a local Monte-Carlo algorithm based on
an auxiliary electric field $\E$. This method also requires an effort
of $O(N+M)$ per sweep and leads to substantial speedups compared to
conventional molecular dynamics methods.  Again the algorithm is
easily parallelizable. Introduction of cluster moves for the field
$\E$ give further increases in efficiency in the $O(M)$ mesh work; this is
most useful when $N/M$ is small.

\section{Particle in Cell codes in Plasma Physics}

Particle in cell methods \cite{buneman,plasmabook} are widely used in
the simulation of plasmas coupled to Maxwell's equations.
Applications are widespread and include astrophysics, laser physics
and fusion research. In these algorithms the continuum magnetic and
electric fields are discretized on mesh. Particles, however, still
move in the continuum:- just like mesh based molecular dynamics
algorithms. The principle technical difficulty lies in correctly
coupling the field and particle degrees of freedom.  The coupling is
usually performed by interpolation of both the {\sl electric charges}
and the {\sl electric currents} to the mesh and then by a back
interpolating the electrodynamic forces from the mesh onto the
particles.

Typically the charge of the particles, $e_i$ is interpolated to the
nodes while the current, $\J_i={\bf v}_i e_i$ is assigned to the
links. This double interpolation is difficult; conservations laws may
not be  preserved.  The importance of exact implementation
of conservation laws is now widely recognized from the field of {\sl
  geometric integration} \cite{geometric}.  Many charge/current
interpolation schemes do not satisfy the continuity equation
\begin{equation}
{\partial \rho\over \partial t} + \div \J=0
\label{continuity}
\end{equation}
We shall show below that the breakdown of continuity leads to errors
in the electrostatic interactions between charges. 

Until now the principle solutions to this problem of charge continuity
were either the use of a low order (and thus noisy) interpolation
scheme for which continuity is exact \cite{buneman} or the use of
higher order schemes with frequent global corrections in which the
longitudinal field components are regularly thrown out and then
reinitialized from solution of the Poisson equation.  A recent and
interesting hybrid approach considers modifications to Maxwell's
equations \cite{plasma,charge} which stabilize their solutions even in
the presence of inconsistencies in the interpolation scheme.

The use of a lattice to interpolate the field degrees of freedom leads
to two other sources of error in the energy of a system of charged
particles in the quasi-static limit.  Firstly the pair interaction
between two particles is given by a lattice Green function, rather
than the exact interaction in $1/r$. In the simplest discretizations
this leads to an interaction potential of the form
\begin{math}
  {1/ r} + O({a ^2/ r^3})
\end{math}
where $a$ is the lattice spacing.  A second and more serious artefact
is the variation of the {\sl self-energy} of the particles as a
function of their position with respect to the lattice. This leads to
a periodic 1-body potential for a particle $i$ with an amplitude which
varies as $e_i^2/\epsilon_0 a$ \cite{rottler1}: As the lattice spacing
decreases this energy variation {\sl diverges} so the discretization
becomes worse.  In the plasma literature this 1-body energy is often
ignored, one simply uses a very coarse mesh with large $a$ so that the
kinetic energy is large compared with this energy scale; in a
thermalized plasma the mesh size must be larger than the Bjerrum
length, $l_b = e^2/4 \pi \epsilon k_B T$.

In mesh based molecular dynamics codes these two artefacts in the
energy are also present. Usually the errors are reduced by convolving
the charge of each particle over a large number of lattice points
\cite{darden,sagui,holm2}, or equivalently by multiplying the
structure factor in Fourier space by an appropriate shaping factor. We
need to implement similar methods if we wish to use particle in cell
techniques for molecular dynamics.

\subsection {Implementation for Molecular Dynamics}

Recently two groups have implemented a molecular dynamics algorithm
based on particle in cell discretizations. As noted above conventional
formulations are not quite good enough for direct application in
molecular dynamics. This new work has 
\begin{itemize}
\item Shown how Maxwell's equations give  Coulomb's law
\item Led to new charge conserving integrators
\item  Controled lattice artefacts in the energy
\end{itemize}
In our own work \cite{joergmd} we used interpolation
algorithms based on polynomial splines in order to 'tune' the error in
the 1-body and pair potentials in a manner very similar to that used
in conventional molecular dynamics codes based on Poisson solvers
\cite{sagui}. To do this we generalized the charge conserving
algorithm of \cite{buneman}, valid for  linear interpolation to 
general B-spline interpolation. The resulting algorithm \cite{joergmd}
conserves charge to machine accuracy for finite time steps and is time
reversible.  In order to simplify the implementation we chose to {\sl
  drop} the Lorentz force from the particle equations of motion which
leads to non-Hamilton (and thus non symplectic) dynamics; however the
dynamics {\sl do} still conserve the total volume in the appropriately
defined configuration space. This conservation law implies the
existence of a Gibbs distribution for the combined field-particle
system.

\begin{figure}[htbp]
  \includegraphics[scale=.55] {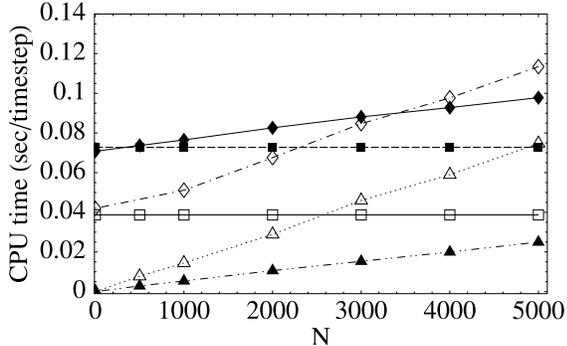}
\caption{\label{fig-linear} Cpu time for the field integration
  $(\square)$, field-particle couplings $(\triangle)$ and total time
  $(\lozenge)$ in a system of size $L=30a$ as a function of $N$ on an
  single AMD Athlon CPU. Also shown are results for the same system
  treated with the P3M method using a charge interpolation of the same
  order (filled symbols), $\blacksquare$ measures the total time in
  Fourier based work. }
\label{fig:relative}
\end{figure}

Our molecular dynamics code has been run in two distinct modes.
Firstly a mode in which the transverse components of the electric
field and the particles are coupled to a thermostat. For this mode we
are able to demonstrate, analytically, convergence to a
Boltzmann-Gibbs distribution of particles interacting as if with a
{\sl instantaneous, non-retarded Coulombic interaction}.  In the
second mode we continue to thermostat the particles but anneal the
transverse components of the electric field to zero temperature.  This
mode is very similar to Car-Parrinello \cite{car} simulations in
quantum molecular dynamics.  Here it is the ratio of the typical
particle velocity, $\bar v$ to the speed of light, $c$ which is tuned
as the optimization parameter.  We expect (but here have no proof)
that the correct Gibbs distribution is generated when $\bar v/c$ is
sufficiently small. We again verified numerically that the correct
effective interaction is found.  A comparison of the speed of our code
compared with a conventional Fourier based Poisson solver
\cite{lammps} is given in Fig.~(\ref{fig:relative}). At very low
densities our code has a slight advantage, at higher densities the
conventional algorithm behaves better. The advantages in each regime
however are rather modest.

Pasichnyk and Duenweg \cite{igor} have taken a rather different and
more direct route to eliminating the errors in the potentials. They
make the remark that for low order interpolation schemes one can
rather easily calculate the exact numerical value of the 1-body
potential.  They then perform a direct, numerical subtraction of this
energy in their code. They have performed a detailed comparison of
this simple and elegant solution with simulations performed with
conventional Fourier methods. They find that their code is competitive
in both speed and accuracy with conventional methods.  The authors
work again in the limit $\bar v/c$ small but with no thermostating of
the electric field. In contrast with our own work they have a full
parallel implementation using MPI.

\section{Origin of Coulomb's law}

In electromagnetism Coulomb's law comes \cite{schwinger} from a local
expression for the energy:
\begin{math}
{ U}= \int {\epsilon_0 \E^2/ 2} \dV 
\end{math},
and the imposition of Gauss' law
\begin{math}
\div \E - \rho/\epsilon_0 =0. 
\end{math}
This motivates the use of the following partition function for the
electric field 
\begin{equation}
 \Z( \rho) = \int {\cal D}\E\,\prod_\r
\delta
\left(
\div  \E - \rho/\epsilon_0\right)
\,e^{- U/k_BT}.
\label{eq-Z}
\end{equation}
We change integration variables to
\begin{math}  
  {\bf e} = { \bf E} +\nabla \phi
\end{math}
with $\nabla^2 \phi= -\rho/\epsilon_0$ and find
\begin{equation}
 {\cal Z} = \int {\cal D}{\bf e}\; \delta(\div {\bf
  e})\, e^{ -\beta {\epsilon_0\over 2} \int ( {\bf e} - \nabla \phi )^2 \dV }.
\end{equation}
The cross term in the exponential is zero by
integration by parts:
\begin{math}
{\int \nabla \phi \cdot {\bf e} \dV }=0
\end{math},
so that
\begin{equation}
 {\cal Z}= e^{-\beta {\epsilon_0 \over 2} \int (\nabla \phi )^2} \int
{\cal D}{\bf e}\; \delta(\div {\bf e}) e^{ -\beta {\epsilon_0\over 2} \int {\bf
    e} ^2 \dV}
\end{equation}
We conclude that $ {\cal Z} = {\cal Z}_{Coul} \times {\rm const}$,
with $Z_{Coul}$ the partition function of particles interacting with
the conventional Coulomb expression: Relative statistical weights are
unchanged if we allow the transverse field to vary freely, rather than
quenching it to zero.  To sample the partition function
Eq.~(\ref{eq-Z}) we have to find solutions to Gauss' law.  Integrating
Maxwell's equations is one method of continuously and locally updating
the system in such a way that Gauss' law is always satisfied.

We  now understand the importance of charge conservation in the
generation of Coulomb's law: Consider the Maxwell equation
\begin{equation}
\epsilon_0 {\partial \E \over \partial t}= -\J + \curl {\bf H}
\label{maxwell}
\end{equation}
and take its divergence, 
\begin{math}
 \div ( \epsilon_0 \dot \E+\J)=0
\end{math}.
If we impose continuity Eq.~(\ref{continuity}) then we conclude that
\begin{equation}
{d\over dt} ( \epsilon_0\, \div \E -\rho)=0
\end{equation}
and Gauss' law is valid. In the absence of continuity Gauss' law is
violated and from the above argument Coulomb's law is not generated as
the effective {\sl thermodynamic} interaction between charges.

\section{Monte-Carlo sampling}

As well as performing molecular dynamic sampling of Eq.~(\ref{eq-Z})
we have performed Monte-Carlo sampling of the constrained partition
function for the case a simple lattice gas \cite{vincent} and also
\cite{rottler1} for off-lattice models. These codes work with a
mixture of two different Monte-Carlo moves. The first is the motion of
a particle together with a slaved update of the electric field in the
vicinity of the particle. The second is a grouped update of the
electric field on the four links defining a plaquette of the mesh. The
two moves are randomly mixed so that the probability of plaquette
updates is $p$. These two moves have been related to the two terms
on the right hand side of Eq.~(\ref{maxwell}) \cite{acm-dynamics}.

\begin{figure}[htbp]
  \includegraphics[scale=.55] {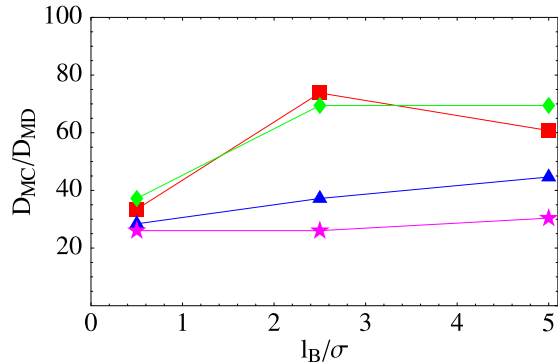}
\caption{Ratio of single particle diffusion coefficients in a simple electrolyte
  obtained from molecular dynamics $D_{MD}$, and Monte Carlo $D_{MC}$;
  time is measured in sweeps for each code. The particles also
  interact with a truncated, repulsive Lennard Jones potential of
  range $r_c=2^{1/6}\sigma$, where $\sigma=2a$.  Measurements as a
  function of relative interaction strength for several
  volume fractions: 
$\Phi=0.15$ $(\blacksquare)$,
 $\Phi=0.3$  $(\blacklozenge)$, 
$\Phi=0.45$ $(\blacktriangle)$, 
$\Phi=0.6 $ ($\bigstar$).
}
\label{fig:mdmc}
\end{figure}

We now compare molecular dynamics and Monte-Carlo codes for
efficiency: {\it a priori} Monte-Carlo should be faster.
\begin{itemize}
\item One does not restrict step sizes in order to ensure accuracy
 and stability of the dynamics.
\item Energy is easier to calculate than force, particularly when
 interpolating to and from a mesh.
\end{itemize}
We explore the first point in Fig.~(\ref{fig:mdmc}) by comparing the
single particle diffusion coefficients in a simple electrolyte as a
function of interaction strength (Bjerrum length) using molecular
dynamics parameters from \cite{holm2}. We also added a truncated,
repulsive Lennard Jones potential to prevent overlap of the particles.
Simulation time is measured in sweeps, and the interaction strength is
varied by changing the Bjerrum length $l_b$ from $\sigma/2$ (weak
Coulomb interaction) to $5 \sigma$ (strong interaction). We find that
in both codes the diffusivity decreases with increasing $l_b$, but the
ratio remains roughly constant. At low volume fractions the Monte
Carlo code is up to 80 times faster than the molecular dynamics code.
This advantage begins to weaken at higher volume fractions, when
collisions (due to the LJ potential) between particles slow down the
diffusion apparently more strongly in MC. Nevertheless Monte Carlo
always retains an advantage. We have not evaluated the second factor
in the relative efficiencies since our two codes are structured very
differently.


Until now we have only considered media with uniform dielectric
properties however Maxwell's equations are valid for arbitrary spatial
variation of $\epsilon(\r)$.  They remain local and thus give an
$O(N)$ algorithm for treating implicit solvents in a situation where
conventional Fourier codes break down.  The Monte-Carlo algorithm can
also be applied to such media \cite{auxiliary}.  To do this one works
with the electric displacement $\D$ rather than the electric field.
The Gauss law constraint now reads
\begin{math}
\div \D = \rho
\end{math},
 the local expression for the energy is  given by
\begin{math}
U= \int 
{\D^2/ 2 \epsilon(\r)}
\dv 
\end{math}.
\subsection{Effective Monte-Carlo Field Dynamics}
Our Monte-Carlo algorithm is based on local updates to both the
particle positions and the electric field.  Both the particles and the
electric field exhibit diffusive dynamics, in contrast to the
propagative behaviour of free particles and waves in the Maxwell
equations. We have shown that the field dynamics are given by the
Langevin equation \cite{acm-dynamics}:
\begin{equation}
  {\partial \E \over \partial t} = 
 D\left  ( \nabla^2 \E - \grad \chargedensity/\epsilon_0  \right)  
-\J/\epsilon_0  +\vec \xi(t)
\label{basic}
\end{equation}
which {\sl replace} Maxwell's equations in our Monte-Carlo scheme.
 Eq.~(\ref{basic}) is particularly
interesting when there are free, mobile charges so that the electric
current is linked to the electric field via the equation $\J=\sigma
\E$. In this case the dispersion law of the transverse electric field
develops a {\sl gap} so that 
\begin{math}
i \omega = \sigma/\epsilon_0 + D q^2
\label{gap}
\end{math}

The relative diffusion rates of the electric and charge degrees of
freedom can be adjusted by modifying $p$ and thus $D$. We
initially believed that the algorithm should be run in a regime in
which the effective diffusion coefficient of the electric field, $D$ is
comparable to or somewhat larger to that of the particles: The
particles are then always interacting via a field that is close to
equilibrium and are not slowed down by the ``drag'' from the electric
degrees of freedom.

Numerical experimentation has shown that one should use a much lower
rate of update.  There are several ways of understanding this
surprising efficiency of the code
\begin{itemize}
\item The gap in the electric dispersion law gives a fast relaxation
  of the electric degrees of freedom {\sl even at very long
    wavelengths}.
\item In certain limits Coulomb interactions are generated even if the
  plaquette updates are suppressed entirely ($p=0)$: a most
  surprising and deep result linked with the statistical mechanics of
  discrete plaquette models \cite{alet}.
\end{itemize}

\subsection{Cluster Monte-Carlo for $\E$}

The efficiency of the Monte-Carlo algorithm is in part due to the gap
which occurs in the dispersion law for the transverse electric field.
This gap has as its origin the finite conductivity of a system with a
finite density of free charges.  What happens when this density is
small or zero? Are we obliged to spend much more of our computer
budget updating the plaquettes in order to increase $D$ and thus
sample the $O(M)$ electric degrees of freedom? For $N/M$  small this
would reduce the effective efficiency of the algorithm. Recently we
have found a way \cite{lucas} of reintroducing a gap in the electric
field spectrum even when $\sigma=0$ based on ideas used to accelerate
the simulation of quantum spin models \cite{alet}.  This quantum
algorithm is in turn inspired by the Swendsen-Wang algorithm for the
Ising model. In the language of electrostatics this new cluster
algorithm works by nucleating a pair of virtual particles of opposite
sign.  These particles then evolve by kinetic Monte-Carlo until they
annihilate taking the system back to a state of no free particles. The
virtual particles re-establish the gap in the electric spectrum and
lead to resampling of the electric field in $O(1)$ sweeps. The
algorithm gives a dynamic exponent for the field $z=0$, rather than
$z=2$ which is implied by Eq.~(\ref{basic}) when $\sigma=0$.

\section{Perspectives}

We have introduced a series of algorithms for the local simulation of
particles interacting via electrostatic interactions. Molecular
dynamics implementations are broadly similar in performance to
conventional Fourier based solvers. They can, however, be trivially
parallelized in an environment with limited interprocessor bandwidth;
this is much harder to do with Fourier based solvers.

We have also formulated an off-lattice Monte-Carlo algorithm that is
faster than the molecular dynamics equivalent while keeping the
advantage of locality.  Finally analogies with ``worm'' or quantum
cluster algorithms allow one to integrate over the field degrees of
freedom with remarkable efficiency.

We note that the above discretizations of electrostatics are very
similar to those introduced in the 19'th century. At this time there
was a very poor intuition as to the physical content and implications
of Maxwell's equations.  FitzGerald introduced a series of mechanical
models of {\sl the ether} which are essentially conserving {\sl
  spatial} discretizations of the continuum Maxwell equations. In
these models a series of wheels are connected by elastic bands
\cite{darrigol}. The rotational motion of the wheels is equivalent to
the magnetic field while the extension, $\E$ of the elastic bands is
equivalent to the electric field. If we perform Monte-Carlo on
FitzGerald's wheels we again recover our local algorithm for
Monte-Carlo simulation.

\bibliography{mc}

\begin{thebibliography}{10}
\expandafter\ifx\csname url\endcsname\relax
  \def\url#1{\texttt{#1}}\fi
\expandafter\ifx\csname urlprefix\endcsname\relax\def\urlprefix{URL }\fi

\bibitem{schlick}
T.~Schlick, R.~D. Skeel, A.~T. Brunger, L.~V. Kal{\'e}, J.~A. {Board, Jr.},
  J.~Hermans, K.~Schulten, Algorithmic challenges in computational molecular
  biophysics, J. Comp. Phys. 151 (1999) 9--48.

\bibitem{darden}
U.~Essmann, L.~Perera, M.~L. Berkowitz, T.~Darden, H.~Lee, L.~G. Pedersen, A
  smooth particle mesh ewald method, J. Chem. Phys. 103~(19) (1995) 8577--8593.

\bibitem{buneman}
J.~Villasenor, O.~Buneman, Rigorous charge conservation for local
  electro-magnetic field solver, Comp. Phys. Comm. 69~(2-3) (1992) 306--316.

\bibitem{plasmabook}
C.~Birdsall, A.~Langdon, Plasma Physics Via Computer Simulation, Institute of
  Physics Publishing, 1991.

\bibitem{geometric}
E.~Hairer, C.~Lubich, G.~Wanner, Geometric Numerical Integration: Structure
  Preserving Algorithms for Ordinary Differential Equations, Springer-Verlag,
  2002.

\bibitem{plasma}
C.-D. Munz, P.~Omnes, R.~Schneider, E.~Sonnendrücker, U.~Voss, Divergence
  correction techniques for maxwell solvers based on a hyperbolic model., J.
  Comput. Phys. 161~(2) (2000) 484--511.

\bibitem{charge}
C.-D. Munz, R.~Schneider, E.~Sonnendrücker, U.~Voss, Maxwell's equations when
  the charge conservation is not satisfied, C. R. Acad. Sci. Paris Sér. I Math.
  328~(5) (1999) 431--436.

\bibitem{rottler1}
J.~Rottler, A.~C. Maggs, A continuum, $o(n)$ monte-carlo algorithm for charged
  particles, J. Chem. Phys. 120~(7) (2004) 3119--3129.

\bibitem{sagui}
C.~Sagui, T.~Darden, Multigrid methods for classical molecular dynamics
  simulations of biomolecules, J. Chem. Phys. 114~(15) (2001) 6578--6591.

\bibitem{holm2}
M.~Deserno, C.~Holm, How to mesh up ewald sums. i. a theoretical and numerical
  comparison of various particle mesh routines, J. Chem. Phys. 109~(18) (1998)
  7678--7693.

\bibitem{joergmd}
J.~Rottler, A.~C. Maggs, Local molecular dynamics with coulombic interaction,
  cond-mat/0312438 .

\bibitem{car}
R.~Car, M.~Parrinello, Unified approach for molecular dynamics and
  density-functional theory, Phys. Rev. Lett. 55~(22) (1985) 2471--2474.

\bibitem{lammps}
S.~J. Plimpton, R.~Pollock, M.~Stevens, Particle-mesh ewald and rrespa for
  parallel molecular dynamics simulations, Proc of the Eighth SIAM Conference
  on Parallel Processing for Scientific Computing .

\bibitem{igor}
I.~Pasichnyk, B.~Duenweg, Coulomb interactions via local dynamics: A
  molecular--dynamics algorithm, cond-mat/0406223 .

\bibitem{schwinger}
J.~S. Schwinger, L.~L. Deraad, K.~A. Milton, W.~yang Tsai, Classical
  Electrodynamics, Perseus Books, 1998.

\bibitem{vincent}
A.~C. Maggs, V.~Rossetto, Local simulation algorithms for coulomb interactions,
  Phys. Rev. Lett. 88~(19) (2002) 196402.

\bibitem{acm-dynamics}
A.~C. Maggs, Dynamics of a local algorithm for simulating coulomb interactions,
  J. Chem. Phys. 117~(5) (2002) 1975--1981.

\bibitem{auxiliary}
A.~C. Maggs, Auxiliary field monte-carlo for charged particles, J. Chem. Phys.
  120~(7) (2004) 3108--3118.

\bibitem{alet}
F.~Alet, E.~S. Sorensen, Cluster monte-carlo algorithm for the quantum rotor
  model, Phys. Rev. E 67~(1) (2003) 015701.

\bibitem{lucas}
L.~Levrel, F.~Alet, J.~Rottler, A.~C. Maggs, Local simulation algorithms for
  coulombic interactions, Proceedings Statphys 22 (2004) .
\newline\urlprefix\url{http:/hogarth.pct.espci.fr/\~{}tony/elec/sp7.pdf}

\bibitem{darrigol}
O.~Darrigol, Electrodynamics from Amp\`ere to Einstein, Oxford University
  Press, 2000.

\end{thebibliography}

\end{document}